\documentclass[journal=nalefd,manuscript=article]{achemso}
\usepackage[version=3]{mhchem}
\usepackage{siunitx}

\author{Shuichi Iwakiri}
\email{siwakiri@phys.ethz.ch}
\author{Folkert K. de Vries}
\author{El\'ias Portolés}	
\author{Giulia Zheng}
\affiliation{Solid State Physics Laboratory, ETH Zurich,~CH-8093~Zurich, Switzerland}
\author{Takashi Taniguchi}
\affiliation{International Center for Materials Nanoarchitectonics,
National Institute for Materials Science,  1-1 Namiki, Tsukuba 305-0044, Japan}
\author{Kenji Watanabe}
\affiliation{Research Center for Functional Materials,
National Institute for Materials Science, 1-1 Namiki, Tsukuba 305-0044, Japan}
\author{Thomas Ihn}
\affiliation{Solid State Physics Laboratory, ETH Zurich,~CH-8093~Zurich, Switzerland}
\alsoaffiliation{Quantum Center, ETH Zurich,~CH-8093 Zurich, Switzerland}
\author{Klaus Ensslin}
\affiliation{Solid State Physics Laboratory, ETH Zurich,~CH-8093~Zurich, Switzerland}
\alsoaffiliation{Quantum Center, ETH Zurich,~CH-8093 Zurich, Switzerland}

\title{Gate-defined electron interferometer in bilayer graphene}

\begin{document}

%

\begin{abstract}
We present an electron interferometer defined purely by electrostatic gating in encapsulated bilayer graphene. This minimizes possible sample degradation introduced by conventional etching methods when preparing quantum devices. The device quality is demonstrated by observing Aharonov-Bohm (AB) oscillations with a period of $h/e$, $h/2e$, $h/3e$, and $h/4e$, witnessing a coherence length of many microns. 
The AB oscillations as well as the type of carriers (electrons or holes) are seamlessly tunable with gating.
The coherence length longer than the ring perimeter and semiclassical trajectory of the carrier are established from the analysis of the temperature and magnetic field dependence of the oscillations.
Our gate-defined ring geometry has the potential to evolve into a platform for exploring correlated quantum states such as superconductivity in interferometers in twisted bilayer graphene.
\end{abstract}

\maketitle
\newpage
There are various ways of tuning the electronic properties of bilayer graphene with electrostatic gating. Its bandgap can be controlled by the displacement field \cite{doi:10.1126/science.1130681,Oostinga2008,PhysRevLett.99.216802,Zhang2009} and this has led to the realization of high-quality quantum devices such as quantum point contacts \cite{Overweg2018,PhysRevLett.107.036602} and quantum dots \cite{Allen2012,doi:10.1021/nl301986q,PhysRevX.8.031023}. Recently, superconductivity has also been discovered in twisted bilayer graphene \cite{Cao2018} and natural (Bernal-stacked) bilayer graphene \cite{zhou2021isospin}.
Superconducting nanodevices in twisted graphene layers have been realized, such as gate-defined Josephson junctions \cite{DeVries2021,Rodan-Legrain2021}, nanowires \cite{Thomson2022}, and SQUIDs \cite{Portoles2022}.

In this work, we introduce the fabrication and investigation of a gate-defined electron interferometer in bilayer graphene, a new member of the gate-defined device family that works as a versatile platform for establishing coherent electron transport via the Aharonov-Bohm effect \cite{PhysRev.123.1511,PhysRev.115.485}. Similar geometries are envisioned for superconducting platforms to resolve the Cooper pair's symmetry in the superconducting regime (Little-Parks effect \cite{PhysRevLett.9.9} and SQUID).
Our method is free of physical or chemical etching of the ring, which has been used conventionally in ring-shaped devices in graphene \cite{Russo2008,Dauber2017,Huefner2010,Yoo2010,Smirnov2012,Schelter2012,Smirnov2014,Damien2017}.
Thanks to this advantage, we could minimize the degradation of the sample quality due to roughness, defects, impurities, or unintended charge localization along edges roughened by etching \cite{Bischoff2015,Bischoff2016,Terres2016}, which has so far set a limit to the advance of such experiments.
Our proof-of-concept device using natural bilayer graphene shows clear Aharonov-Bohm oscillations \cite{PhysRev.123.1511,PhysRev.115.485} with a flux period of $h/e$ and its higher harmonics ($h/2e$, $h/3e$, and $h/4e$). The oscillations can be completely turned on and off through gating. We also study the oscillations as a function of carrier density and temperature and subsequently analyze the behavior of the coherence length extracted from them. Furthermore, we observe limitations of our finite-size geometry through semiclassical effects at a higher magnetic field.
Our device structure is expected to work for various systems such as natural bilayer graphene, twisted bilayer graphene, and heterostructures of bilayer graphene and other van der Waals materials.

\begin{figure}[t]
\centering
\includegraphics[scale=0.95]{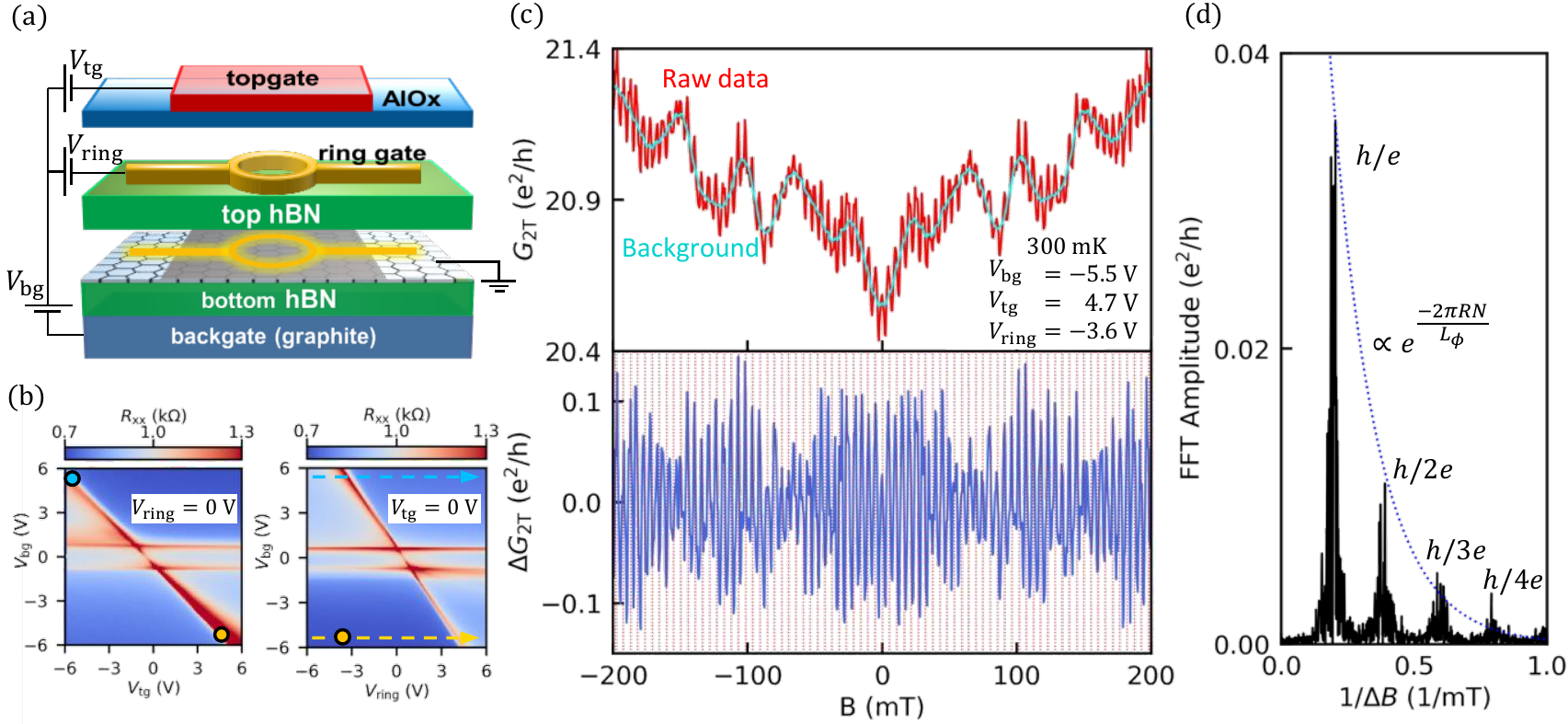}
\caption{(a) Schematic of the sample. Bilayer graphene is encapsulated by two hBN flakes. A graphite backgate, metallic ring-shaped gate (ring gate), and topgate are fabricated. The dark shaded region is where the bilayer graphene is set at the charge neutrality and gapped, while the yellow region has a finite charge density. (b) Resistance map as a function of (left) $V_{\rm bg}$ and $V_{\rm tg}$ with $V_{\rm ring}=0$ V. (right) $V_{\rm bg}$ and $V_{\rm ring}$ with $V_{\rm tg}=0$ V.The condition at which the data in (c) is marked with the yellow circle. The sweep direction for the measurement in Fig. \ref{gatetunability}(a) is shown with yellow and blue dotted arrows. (c) Conductance as a function of the magnetic field with $(V_{\rm bg}, V_{\rm tg}, V_{\rm ring})=$ (-5.5 V, 4.7 V, -3.6 V). Upper panel: Raw data (red) and the smoothed background (cyan). Lower panel: Data after subtraction of the background. Equidistant lines in a period of 5 mT are shown. (d) FFT spectrum of the data shown in the lower panel of (c). Peaks of the harmonics are labeled ($h/e,h/2e,h/3e$ and $h/4e$). The blue dotted line shows the exponential fitting of the 4 peaks.
}
\label{sampleandresult}
\end{figure}

Figure \ref{sampleandresult}(a) shows the schematic of the sample. A stack of hexagonal boron-nitride (top hBN)/bilayer graphene/hBN (bottom hBN)/graphite (back gate) is made.
Electric contacts to the edge of the bilayer graphene are fabricated by etching the top hBN and subsequently depositing Cr/Au. A ring-shaped gate (ring gate) is patterned on the top hBN. The lithographic width of the ring gate is 80 nm and the inner (outer) radius is 500 nm (580 nm). The fundamental period of the AB oscillations is thus expected to be between $\frac{e}{h\times\pi(500\:\rm nm)^2}\sim$ 5.3 mT (for the inner radius) and $\frac{e}{h\times\pi(580\:\rm nm)^2}\sim$ 3.9 mT (for the outer radius).
This is of the same order as the dimensions reported previously in etched graphene rings \cite{Russo2008,Dauber2017,Huefner2010,Yoo2010,Smirnov2012,Schelter2012,Smirnov2014,Damien2017}. The actual electrostatic width of the ring is expected to be larger than the lithographic width (80 nm) due to fringe-field effects since the ring gate is closer to the graphene than the top gate. With an electrostatic simulation, it is estimated to be 160 nm (see Supporting Information). After fabricating the ring gate, an aluminum oxide layer is formed by atomic layer deposition, and the top gate of Cr/Au is deposited covering the area between the contacts. The key feature of this device is that one can independently control the carrier density and displacement field (bandgap) both under the ring gate and elsewhere. This is because the metallic ring gate screens the electric potential of the top gate above it.

We first measure the four-terminal conductance $G_{\rm 4T}$ as a function of carrier density $n$ without defining the ring (not shown). Through this measurement, we estimate the mobility $\mu\sim4\times10^{5}$ cm$^2$/Vs for electrons ($n>0$) and $\mu\sim2.5\times10^{5}$ cm$^2$/Vs for holes ($n<0$), and the mean free path $l_{\rm mfp}$ between 2.7 $\mu$m (at $n=0.4\times10^{12}$  /cm$^2$) to 15 $\mu$m (at $n=2.4\times10^{12}$  /cm$^2$).

To find the condition to form a ring, we measure the two-terminal conductance sweeping the voltages of backgate ($V_{\rm bg}$), ring gate ($V_{\rm ring}$), and topgate ($V_{\rm top}$) within the range $\pm 6$ V. All measurements are performed at 300 mK unless stated otherwise. The left panel of Figure~\ref{sampleandresult}(b) shows the resistance map of $V_{\rm bg}$ versus $V_{\rm tg}$ at $V_{\rm ring}=0$ V. The diagonal line of enhanced resistance between ($V_{\rm bg}$, $V_{\rm tg}$)=(-6 V, 6 V) and (6 V, -5.3 V) is the charge neutrality line under the top gate, with which we form confinement to define a ring. This condition does not depend on $V_{\rm ring}$.
Hereafter, we refer to the region tunable by the topgate as bulk. The two horizontal lines at $V_{\rm bg}=$ 0.91 V and -0.6 V are the charge neutrality lines of regions below the ring gate and the area covered by neither the topgate nor the ring gate, respectively. 
The right panel of Figure~\ref{sampleandresult}(b) shows the resistance map of $V_{\rm bg}$ versus $V_{\rm ring}$ at $V_{\rm tg}=0$ V. The similar behavior to the previous map is observed. We use the parallel-plate capacitor model to obtain the carrier density $n_{\rm ring}$ and the displacement field $D_{\rm ring}$ below the ring-gate and the corresponding bulk quantities $n_{\rm bulk}$ and $D_{\rm bulk}$ (see Supporting Information).
We also take a resistance map by sweeping $V_{\rm bg}$, $V_{\rm tg}$, and $V_{\rm ring}$ (see Fig. \ref{gategatemap}(b) in the Supporting Information), where $V_{\rm ring}$ is swept such that it stays at the charge neutrality condition. By this measurement, we measure the highest achievable resistance of 3.1 kOhm between the two contacts across the ring. 
If the insulation under the ring- and top-gated region is perfect, we would expect a high resistance of the order of MOhm\cite{Overweg2018}. This discrepancy suggests the existence of current leakage paths, which add a background conductance on top of the transport through the ring.

We choose to form the ring with $(V_{\rm bg}, V_{\rm tg}, V_{\rm ring})=$ (-5.5 V, 4.7 V, -3.6 V), where the bulk is gapped and at charge neutrality with $n_{\rm bulk}=0$ and $D_{\rm bulk}=-0.34$ V/nm (yellow circle in Fig. \ref{sampleandresult}(b) left panel), while the ring is conducting with $n_{\rm ring}=-3.6\times 10^{12}$/cm$^2$ and $D_{\rm ring}=-0.38$ V/nm (yellow circle in Fig. \ref{sampleandresult}(b) right panel). The magneto-conductance trace under this condition is shown in Figure \ref{sampleandresult}(c).
The magnetic field is swept in the range $\pm$200 mT in  steps of 500 $\mu$T while measuring the two-terminal conductance $G_{\rm 2T}$ by applying an AC voltage of 10 $\mu$Vrms at 37 Hz.
The conductance at zero magnetic fields $G_{0}$ is of the order of 20 $e^2/h$, which suggests that the number of modes in the ring is of the order of 10. This agrees with the estimation using the width of the ring (160 nm) and the Fermi wavelength $\lambda$ at a given carrier density and the four-fold degeneracy ($\lambda=\sqrt{\frac{4\pi}{|n_{\rm ring}|}}\sim18$ nm).

In the upper panel of Fig. \ref{sampleandresult}(c), one can see periodic conductance oscillations (red line) on top of a smooth background (cyan line). We attribute this background to the universal conductance fluctuations of the conductive part of the sample and we filter them out by numerically smoothening the trace with a Savitzky–Golay filter \cite{SavGol} (window size 40 mT and polynomial order 2). 
By subtracting this smooth background from the raw data, we obtain the oscillatory part of the conductance $\Delta G_{\rm 2T}$ as shown in the lower panel of Fig. \ref{sampleandresult}(c). The period ($\sim$ 5 mT) corresponds to the expectation for AB oscillations. All the results and following discussions are robust with respect to the details of the Savitzky–Golay filtering such as the window size (from 20 to 50 mT) and polynomial order (from 2 to 4).

The root-mean-square of $\Delta G_{\rm 2T}$, namely $\Delta G_{\rm RMS}$, has been used as a measure of the oscillation amplitude\cite{Russo2008,Dauber2017}. The maximum $\Delta G_{\rm RMS}$ of our sample is $\sim 4\times10^{-2}$ $e^2/h$ at 300 mK, which is higher than the previous reports in graphene such as $\sim 9\times10^{-3}$ $e^2/h$ at 150 mK \cite{Russo2008} and $\sim 2\times10^{-2}$ $e^2/h$ at 36 mK \cite{Dauber2017}. If we compared the ratio of $\Delta G_{\rm RMS}$ and the background conductance at zero magnetic field $G_{0}$, the value of $\Delta G_{\rm RMS}/G_{0}$ of our sample (0.2 $\%$) is comparable to previous reports 0.15 $\%$ \cite{Russo2008} and 0.34 $\%$ \cite{Dauber2017}.

We perform a fast Fourier transform (FFT) of $\Delta G_{\rm 2T}$ resulting in the spectrum shown in Fig. \ref{sampleandresult}(d). The FFT spectrum shows a pronounced peak at around 0.2 mT$^{-1}$ (period 5 mT) corresponding to the flux quantum $h/e$ through the ring area with an effective ring radius $R\simeq513$ nm. The FFT spectrum also shows peaks at 0.4 ($h/2e$), 0.6 ($h/3e$), and 0.8 ($h/4e$) mT$^{-1}$, which are the harmonics of $h/e$ (each flux quantum shown in the parenthesis). 
The $N$-th harmonic originates from pairs of clockwise and counterclockwise paths in the ring with a winding number difference $N$. The observation of higher harmonics suggests that the coherence length is at least comparable to the ring perimeter.
This is an improvement in comparison to  conventional etching-defined rings. Previously,  oscillation periods up to $h/e$ \cite{Huefner2010,Smirnov2012,Smirnov2014,Damien2017} or $h/2e$ \cite{Russo2008} have been observed and $h/3e$ oscillations were reported only in an etched sample encapsulated by hBN measured at a temperature of 36 mK \cite{Dauber2017} which is more than a factor of 8 lower than our temperature (300 mK). Here, the observation of the $h/4e$ harmonic indicates that the coherence length is larger than in samples prepared by etching. We attribute this improvement to the use of the gate-defined geometry in combination with the hBN-encapsulation technique and the use of an exfoliated graphite backgate, which is expected to be atomically flat.
Thanks to the observation of these higher harmonics, one can perform a more detailed analysis of the data. The relative amplitude of the $N$-th harmonic is expected to decay exponentially $\propto \rm exp(-2\it \pi RN/L_{\phi})$ \cite{Hansen2001} with $L_{\phi}$ the phase coherence length. If we extract $L_{\phi}$ by fitting the peaks in the FFT spectrum (Fig. \ref{sampleandresult}(d)), we obtain $L_{\phi}\simeq 2.8 \pm 0.2$ $\mu$m. We repeat the same calculation for different $n_{\rm ring}$, and obtained the maximum value of $L_{\phi}\sim4.4\pm0.5$ $\mu$m at $n_{\rm ring}=3.94\times10^{12}$ /cm$^2$. This result is consistent with the hypothesis that phase coherence is well maintained on the scale of the ring perimeter $2\pi R\simeq\SI{3.1}{\mu m}$. The above results demonstrate the realization of a high-quality quantum interferometer with a gate-defined technique.

\begin{figure}[H]
\begin{center}
\includegraphics[scale=1,pagebox=cropbox,clip]{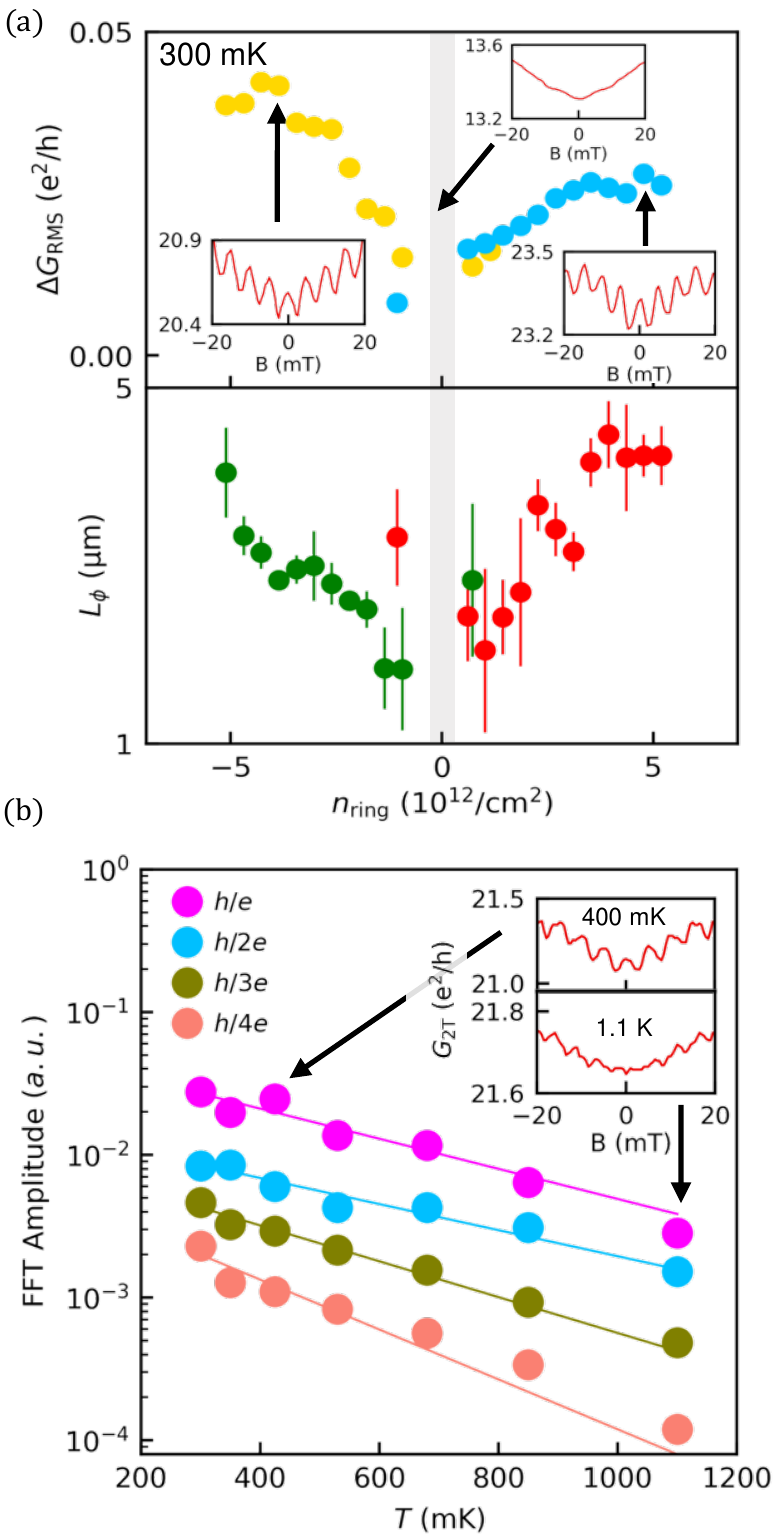}
\caption{(a) Gate tunability of the AB oscillations. The same values as Fig. \ref{sampleandresult}(b) are used for $V_{\rm bg}$ and $V_{\rm tg}$. Upper panel: Oscillation amplitude $\Delta G_{\rm RMS}$ vs $n_{\rm ring}$. The inset shows the raw data at three points indicated by arrows. Lower panel: Phase coherence length $L_{\phi}$ vs $n_{\rm ring}$. The point around the charge neutrality is gray-shaded since AB oscillations are not observed.  (b) Temperature dependence of the AB oscillations, evaluated by the FFT peak value of each harmonic ($h/e, h/2e, h/3e,$ and $h/4e$). $V_{\rm bg}$, $V_{\rm tg}$, and $V_{\rm ring}$ are the same as Fig. \ref{sampleandresult}(b). Solid lines are the fit to the $\propto \rm exp\it (-T/T_{\rm 0})$. The insets show the raw data at 400 mK and 1.1 K.}
\label{gatetunability}
\end{center}
\end{figure}

Furthermore, the observed AB oscillations are tunable by gating. We measure the $n_{\rm ring}$ dependence of the AB oscillations by sweeping $V_{\rm ring}$ along with the dotted arrows in Fig. \ref{sampleandresult}(b). In this measurement, $V_{\rm bg}$ and $V_{\rm top}$ are fixed to keep the bulk properties constant ($n_{\rm bulk}=0$ and $D_{\rm bulk}=-0.34$ V/nm), whereas $V_{\rm ring}$ is swept. 
We use the root-mean-square conductance $\Delta G_{\rm RMS}$ after subtracting the background from the magnetic-field dependent conductance as a measure of the oscillation amplitude as shown in Fig. \ref{gatetunability}(a).
The insets show the raw data at $n_{\rm ring}=$-3.86, 0, and 4.77 $\times10^{12}$/cm$^2$. The oscillations are clearly seen in both the electron and hole regions. The gray-shaded region at around $n_{\rm ring}\sim 0$ is where the oscillation is no longer observed.
The oscillation amplitude depends on $n_{\rm ring}$, changing from $4.0\times10^{-2}$ to $1.6\times10^{-3}$ $e^2/h$ with decreasing density. 
%
When we sweep $V_{\rm ring}$ fixing $V_{\rm bg}$ and $V_{\rm tg}$, we are essentially sweeping both the carrier density $n_{\rm ring}$ and the displacement field $D_{\rm ring}$ at the same time. To evaluate the contribution of them separately, We have also performed sweeps of $n_{\rm ring}$ with $D_{\rm ring}$ fixed at different values (and vice versa). We find that $n_{\rm ring}$ is the dominant factor determining the oscillation amplitude (see Supporting Information).

The lower panel of Fig. \ref{gatetunability}(a) shows the $n_{\rm ring}$ dependence of $L_{\phi}$ which has the same tendency as that of $\Delta G_{\rm RMS}$, showing a decreasing trend towards $n_{\rm ring}\rightarrow 0$. This behavior can be primarily attributed to the carrier density dependence of the Fermi velocity $v_{\rm F}$. The $v_{\rm F}$ is the group velocity of electrons passing through the ring, and it is a monotonic function of the carrier density (e.g. $v_{\rm F}=\hbar \sqrt{\pi n_{\rm ring}}/m$ in bilayer graphene. $m$ is the effective mass). Then, $n_{\rm ring}\rightarrow 0$ leads to $v_{\rm F} \rightarrow 0$. As a consequence, the traveling time of the electrons becomes longer, making them subject to stronger decoherence.

Figure \ref{gatetunability}(b) shows the temperature dependence of the FFT amplitude of $h/e$, $h/2e$, $h/3e$, and $h/4e$ peaks from 300 mK to 1.1 K. The carrier density is set at $n_{\rm ring}=-3.6\times 10^{12}$ /cm$^2$ of Fig. \ref{gatetunability}(a) (indicated by arrow). We observe that all harmonics decay approximately exponentially with temperature ($\propto$ exp$(-T/T_{0})$) with a characteristic temperature $T_{0}\sim 360$ mK. 
Recalling that the oscillation amplitude is proportional to $\rm exp(-2\it \pi RN/L_{\phi})$, the exponential temperature dependence suggests that $L_{\phi}\propto 1/T$. Such behavior is expected for a ballistic ring \cite{Seelig2001,Hansen2001}.
In addition, the slope of the temperature dependence slightly differs between harmonic components $N$. Most simply, one expects that the exponent increases with the harmonic number $N$ since the amplitude of the oscillation is proportional to $\rm exp(-2\it \pi RN/L_{\phi})$. In Fig. \ref{gatetunability}(b) or in other experiments in GaAs 2DEG \cite{Hansen2001}, one cannot directly see such clear dependence on $N$ (see $h/e$ and $h/2e$ peaks, for example). The reason is attributed to the different origins of these peaks. Electrons relevant for transport inside the ring have a finite energy variation within about $4k_{\rm B}T$ around the Fermi energy. This varies the relevant wavelength of the electron and thereby the phase that the electron acquires upon traveling, which suppresses the $h/e$ (and $n$-odd) oscillations. However, $h/2e$ (or $n$-even) harmonics are partially due to the interference between time-reversed trajectories, in which the phase accumulated via path length drops out due to time-reversal symmetry \cite{Hansen2001}. This makes the slope of $h/2e$ as a function of temperature less steep than twice that of $h/e$.
One can also estimate the phase coherence length from the temperature dependence of the $h/2e$ peak. If $h/2e$ peak depend on temperature such that $\propto e^{-sT}$ ($s$: constant), the phase coherence length is given by $L_{\phi}=\frac{4\pi R}{sT}$. If we applied this analysis to the data in Fig. \ref{gatetunability}(a) ($T=300$ mK, $n_{\rm ring}=4\times10^{12}$ /cm$^2$, and $s=-2.03\times10^{-3}$ /K), we obtain $L_{\phi}\simeq 10$ $\mu$m. This estimation gives a longer value compared to the exponential fit used in Fig. \ref{sampleandresult}(d) and \ref{gatetunability}(b), since it separates decoherence by interactions from thermal averaging \cite{Hansen2001}.

\begin{figure}[H]
\centering
\includegraphics[scale=1,pagebox=cropbox,clip]{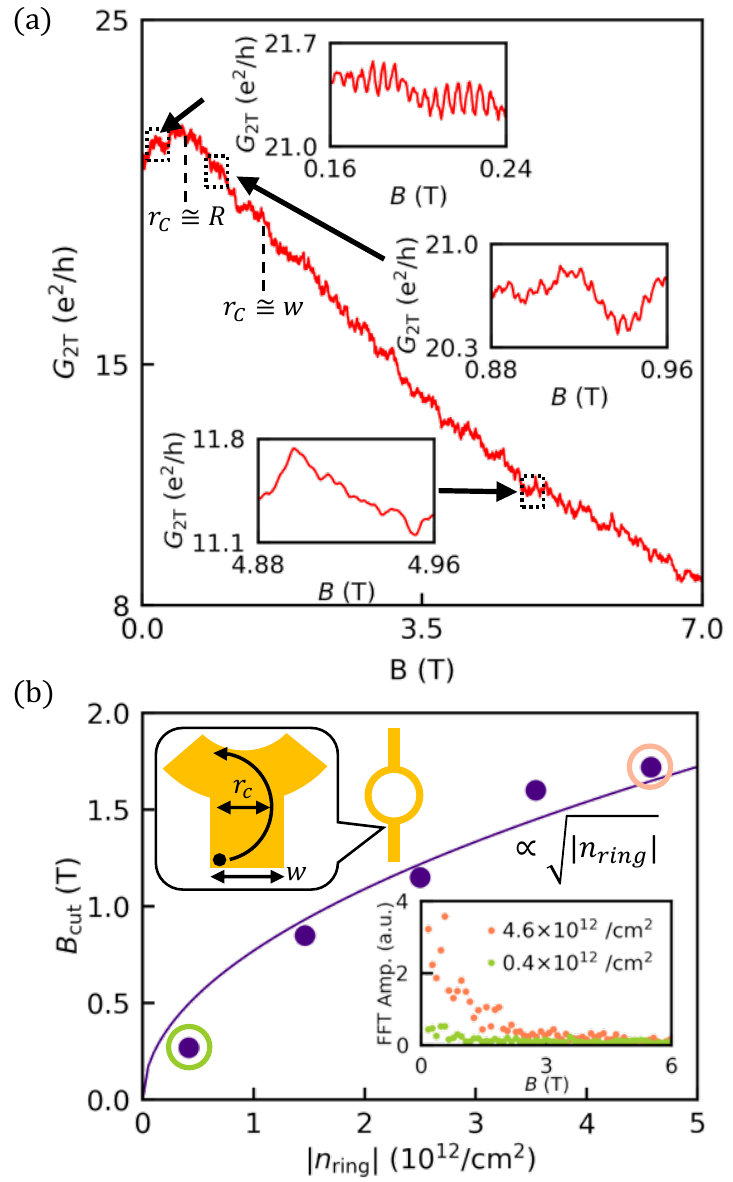}
\caption{(a) Conductance measurement up to 7 T in steps of 500 $\mu$T. $V_{\rm bg}$, $V_{\rm tg}$, and $V_{\rm ring}$ are the same as in Fig. \ref{sampleandresult}(b). The insets show the raw data at magnetic fields of 0.16-0.24 T, 0.88-0.96 T, and 4.88-4.96 T. The points at which the cyclotron radius coincides with the radius ($R$) and the width ($w$) of the ring are indicated.(b) Carrier density $|n_{\rm ring}|$ dependence of the cutoff field $B_{\rm cut}$. $V_{\rm bg}$ and $V_{\rm tg}$ are fixed to the same value as (a). $V_{\rm ring}$ is swept from -5.5, -3.5, -1.5, 0.5, to 2.5 V ($n_{\rm ring}=$ -4.6, -3.5, -2.5, -1.5, and -0.4 $\times10^{12}$/cm$^2$). The solid curve is calculated from the cyclotron radius of the ring width $w=$160 nm. The schematic illustrates the trajectory of the electron's cyclotron motion. The inset shows the magnetic field dependence of the FFT amplitude at $h/e$ peak.}
\label{highmag}
\end{figure}

We also track the behavior of the magneto-conductance up to 7 T as shown in Fig. \ref{highmag}(a). The gate voltages $V_{\rm bg}$, $V_{\rm tg}$, and $V_{\rm ring}$ are the same as in Fig. \ref{sampleandresult}(c). We perform FFTs from finite magnetic field intervals by moving the range in steps of 80 mT to track the development of the oscillation amplitude of the fundamental oscillation as a function of the magnetic field. Fig. \ref{highmag}(a) shows the result of the measurement. The conductance decreases by more than a factor of two, as the field is swept up to 7 T. The insets show the zoom-in picture of each magnetic field region, showing that the oscillation amplitude decreases with increasing magnetic field.
As the magnetic field increases, characteristic length in the system such as the cyclotron radius $r_{\rm c}=\frac{\hbar k_{F}}{eB}=\frac{\hbar \sqrt{\pi |n_{\rm ring}|}}{eB}$ ($e$ is the electron charge) becomes comparable to the dimension of the ring (radius $R$ and width $w$, for example). The dashed lines in Fig. \ref{highmag}(a) show the point at which these length scales match.

We track the FFT amplitude of the $h/e$ peak as a function of the magnetic field (see the inset of Fig. \ref{highmag}(b)). It is observed that the FFT amplitude decreases with the magnetic field. To characterize this, we define a cutoff field $B_{\rm cut}$ at which the FFT peak height becomes 10 $\%$ of its maximum. The data for all the measured $n_{\rm ring}$ is shown in Supporting Information.
Figure \ref{highmag}(b) shows the $B_{\rm cut}$ as a function of $|n_{\rm ring}|$. There is a clear $|n_{\rm ring}|$ dependence which follows a fit as $B_{\rm cut}\propto\sqrt{|n_{\rm ring}|}$.
This dependence suggests the relevance of density-dependent length scales such as the cyclotron radius.
One possible explanation is a semi-classical effect, which becomes prominent when the mean free path of the electron in the ring is longer than (at least) the width of the ring.
Because of the Lorenz force, electrons preferentially enter one arm of the ring which breaks the left-right symmetry necessary for the observation of AB oscillations. When the cyclotron radius $r_{\rm c}$ becomes comparable to the ring width $w$, all incident electrons enter only one of the arms as illustrated in Fig. \ref{highmag}(b). This effect of Lorenz force-induced asymmetry suppresses the electron interference\cite{Szafran2005}.
Now, assuming that the $B_{\rm cut}$ is the magnetic field at which the cyclotron radius $r_{\rm c}$ becomes equal to $w$, we obtain $B_{\rm cut}=\frac{\hbar \sqrt{\pi |n_{\rm ring}|}}{ew}\propto \sqrt{|n_{\rm ring}|}$. This estimation agrees with the data shown in Fig. \ref{highmag}(b) with 
$w = \SI{160}{nm}$ obtained from the sample geometry and electrostatic simulation.

In conclusion, we have presented the fabrication of a high-quality and tunable electron interferometer in bilayer graphene defined purely by electrostatic gating. The quality of the device was demonstrated by the observation of Aharonov-Bohm (AB) oscillations with a fundamental period of $h/e$ and its harmonics up to $h/4e$, from which we deduced a phase coherence length up to $\SI{4.4}{\mu\rm m}$. We also showed the ambipolar operation of the device and the complete tuning of the oscillations varying the carrier density. The temperature and magnetic field-dependent measurements established a long coherence length and semiclassical trajectory under a finite magnetic field. Our gate-defined ring geometry has the potential to work as a platform for probing quantum interference in various systems such as natural bilayer graphene, twisted bilayer graphene, and heterostructures of bilayer graphene and other van der Waals materials.


\begin{acknowledgement}
We are grateful for the technical support from Peter Maerki, Thomas Baehler, and the ETH FIRST cleanroom facility staff. 

We acknowledge financial support from the European Graphene Flagship, the ERC Synergy Grant Quantropy, and the European Union’s Horizon 2020 research and innovation program under grant agreement number 862660/QUANTUM E LEAPS and NCCR QSIT (Swiss National Science Foundation) and under the Marie Sklodowska-Curie Grant Agreement Number 766025. K.W. and T.T. acknowledge support from JSPS KAKENHI (Grant Numbers 19H05790, 20H00354, and 21H05233).
\end{acknowledgement}

\bibliography{MTJ_cite}

\pagebreak
\begin{center}
\textbf{\Large Supporting Information: Gate-defined electron interferometer in bilayer graphene}
\end{center}

\setcounter{section}{0} 
\renewcommand\thesection{Appendix~\Alph{section}} 
\renewcommand\thefigure{S\arabic{figure}} 
\setcounter{figure}{0} 

\subsection{SAMPLE FABRICATION}

\begin{figure}[H]
\centering
\includegraphics[scale=1.5,pagebox=cropbox,clip]{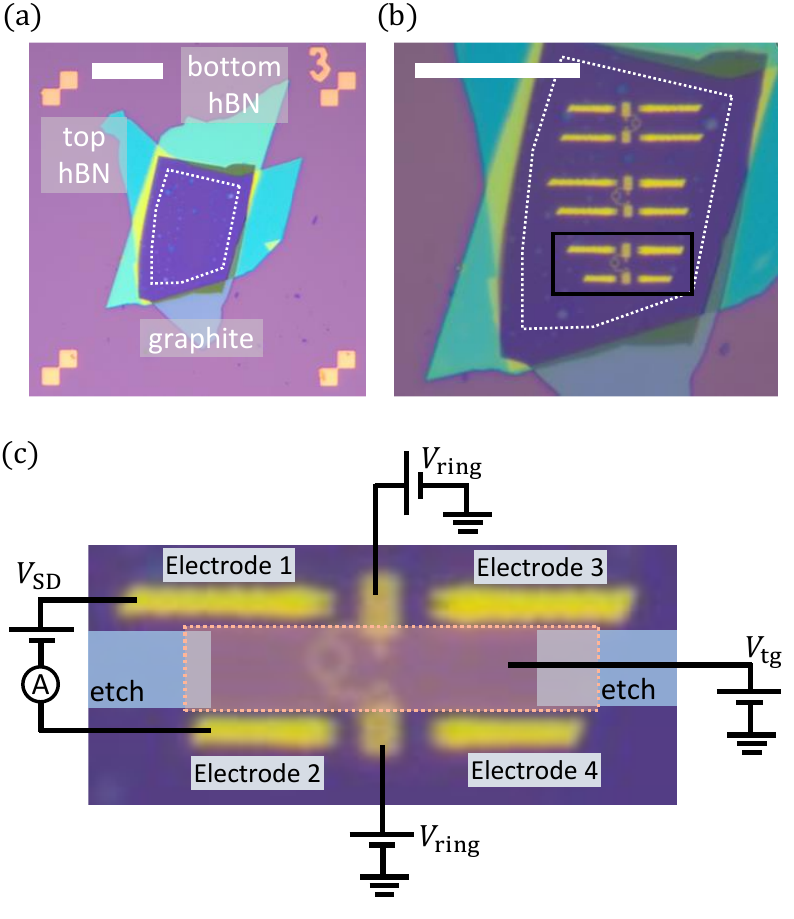}
\caption{Optical micrograph of the (a) whole stack before fabrication. The white dotted area shows the bilayer graphene. Top hBN, bottom hBN, and graphite are indicated. (b) Zoom-in to the fabricated area after making the electrode contact pad and ring gate for three sets of devices. The black rectangle is the area shown in (c). The white scale bar is 20 $\mu$m for both (a) and (b). (c)Sample micrograph with a circuit schematic. The red shaded area is the topgate and the blue shaded area is etched. These pictures are taken before fabricating the lead that connects the pads to the pad for wire bonding on the chip.}
\label{samplepicture}
\end{figure}

\hspace{\parindent} Our sample consists of a multi-stack of hBN (37.3 nm)/bilayer graphene/hBN (45.1 nm)/graphite (30 nm) and deposited on a 285 nm $\mathrm{p:Si/SiO_2}$ substrate using a polymer-based dry transfer technique. For the pick-up phase, we use a self-made polydimethylsiloxane-polycarbonate stamp. The bilayer graphene flake of 30 $\mu$m $\times$ 30 $\mu$m is obtained by mechanically exfoliating a bulk graphite crystal in an ambient condition. The micrograph of the finished stack is shown in Fig. \ref{samplepicture}(a). We employ graphite as the bottom gate and used hBN as the gate dielectric (with dielectric constant $\sim$3.3).

Figure \ref{samplepicture}(b) shows the optical micrograph of the sample after fabricating the contact pad and the ring gate. Electric edge contacts to the bilayer graphene (Electrode 1,2,3, and 4) are made by etching the top hBN (37.3 nm) and half of the bottom hBN (45.1$\times$0.5$\simeq$22.5 nm) in the shape of the electrodes. This is achieved by means of reactive ion etching (RIE) ($\mathrm{CHF_3/O_2}$, 40/4 sccm, 60W, with a hBN etching rate of 0.6 nm/s) and subsequent Cr/Au (10 nm/110 nm) deposition.
A ring-shaped gate (ring gate) of Cr/Au (10 nm/40 nm) is patterned on the top hBN. After fabricating the ring gate, the whole stack is etched to define the mesa and a 30 nm atomic-layer-deposited aluminum oxide (AlOx; dielectric constant $\sim$9.5) is formed. The etching in this process is a few microns away from the ring part and does not affect the performance of the AB oscillations. On top of this AlOx layer, the top gate of Cr/Au (10 nm/110 nm) is patterned to cover the area between the contacts. As shown in Fig. \ref{samplepicture}(b), three sets of devices are made in the stack. The measurements on this paper are performed on the bottom one.
Figure  \ref{samplepicture}(c) shows the zoom-in to the measured sample with circuit schematics.

In the measurement, we use Electrode 1 and 2 are used for the two-terminal conductance measurement by applying a voltage $V_{\rm SD}$ and measuring the current using a home-made IV-converter. Electrode 3 and 4 are used as voltage probes when measuring in the four-terminal configuration or floated otherwise. The topgate is formed in the red shaded area (on top of the AlOx layer) and the blue shaded area is etched to define a mesa, isolating the three devices electrically and removing impurities introduced during the fabrication process such as resist residues.

\subsection{ELECTROSTATIC SIMULATION}
\begin{figure}[H]
\centering
\includegraphics[scale=1,pagebox=cropbox,clip]{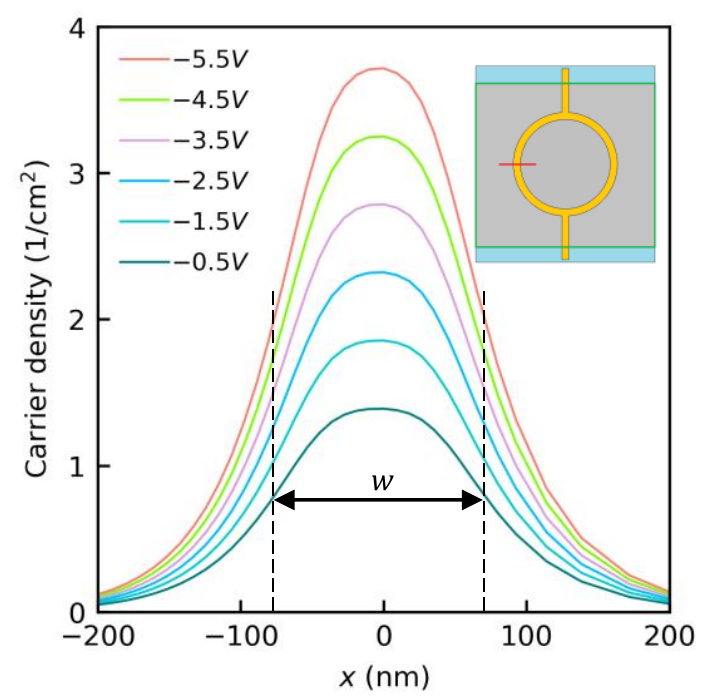}
\caption{Simulated carrier density distribution around the ring gate for $V_{\rm ring}=$-5.5, -4.5, -3.5, -2.5, -1.5, and -0.5 V. The inset shows the top view of the geometry used for the simulation. The yellow (gray) region is the graphene under the ring gate (topgate). The Blue region is the rest of the graphene. Redline shows the line along which the carrier densities are plotted. The electrostatic width of the ring $w$ defined as the FWHM of the peak is also shown.}
\label{simulation}
\end{figure}

To see the effect of gating, we perform electrostatic simulations using COMSOL MULTIPHYSICS software. We input all the geometrical variables such as the dimension of the electrodes and the stack, and we set the $V_{\rm bg}=-5$ V and $V_{\rm tg}=7$ V such that $n_{\rm bulk}=0$. This condition is shifted from the one obtained experimentally ($V_{\rm bg}=-5$ V and $V_{\rm tg}=4.7$ V) because of the discrepancy in the capacitance from the model to the one in the real device. We then calculate the electric potential using the finite-element method and convert the result to the carrier density, sweeping $V_{\rm ring}$.
Figure \ref{simulation} shows the result of the simulation around the ring gate. One can see that the carrier density has a broadened peak shape. We define the full-width-half-maximum (FWHM) of this peak as an electrostatic width of the ring, which is 160 nm for all the $V_{\rm ring}$ values. Note that the lithographic width of the ring gate is 80 nm. This discrepancy is due to the fact that the distance between the graphene and the ring gate (37.3 nm, same as the thickness of the top hBN) is the same order as the lithographic width of the ring gate, which makes the fringing field non-negligible.

\subsection{SAMPLE CHARACTERIZATION}
\begin{figure}[H]
\centering
\includegraphics[scale=1.5,pagebox=cropbox,clip]{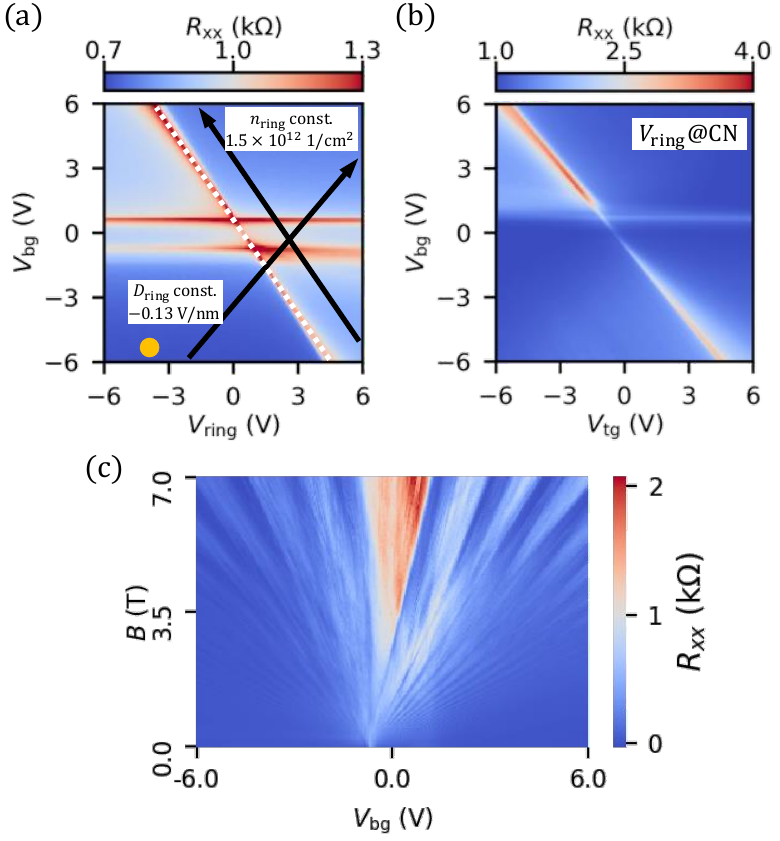}
\caption{Resistance map as functions of (a) $V_{\rm bg}$ and $V_{\rm ring}$ and (b) $V_{\rm bg}$ and $V_{\rm tg}$, keeping $V_{\rm ring}$ at the charge neutrality (CN) line. (c) Two terminal resistance $R_{xx}$ as functions of magnetic field $B$ and backgate voltage $V_{\rm bg}$.}
\label{gategatemap}
\end{figure}

We characterized the sample by measuring the two-terminal resistance sweeping gate voltages under a zero magnetic field. All measurements are done at 300 mK.
Figure \ref{gategatemap}(a) shows the map of the resistance as functions of backgate voltage ($V_{\rm bg}$) and ring gate voltage ($V_{\rm ring}$). Here, $V_{\rm top}$ is fixed at 0 V. A tilted line (shown as white dotted line) from ($V_{\rm bg}$, $V_{\rm ring}$)=(-6 V,4.23 V) and ($V_{\rm bg}$, $V_{\rm ring}$)=(6 V,-5.1 V) corresponds to the charge neutrality (zero-density line) of the area under the ring. Horizontal lines crossing $V_{\rm bg}=$0.91 and -0.6 V are the charge neutrality of the part which is not covered by the ring gate and elsewhere (the region which is not covered either ring gate or top gate), respectively.

We also measure a resistance map by sweeping $V_{\rm bg}$, $V_{\rm tg}$, and $V_{\rm ring}$ as shown in  Fig. \ref{gategatemap}(b). Here, $V_{\rm ring}$ is swept such that it stays at the charge neutrality condition (white dotted line in Fig. \ref{gategatemap}(a)). By this measurement, we measure the highest achievable resistance of 3.1 kOhm between the two electrodes (Electrode 1 and 2 in Fig. \ref{samplepicture}(c)) across the ring at ($V_{\rm bg}$, $V_{\rm tg}$, $V_{\rm ring}$)= (2.5 V, -3 V, -2.4 V).
%

The electrostatics of the stack can be modeled as a parallel place capacitor by using the capacitance between graphene and backgate ($C_{\rm bg}$), graphene and ringate ($C_{\rm ring}$), and graphene and topgate ($C_{\rm top}$).
According to the capacitor model, one can express the carrier density inside the ring ($n_{\rm ring}$) and outside the ring ($n_{\rm bulk}$) as
$n_{\rm ring}=\frac{\varepsilon_{0}\varepsilon_{\rm hBN}}{ed_{\rm bot}}(V_{\rm bg}-V^{0}_{\rm bg})+\frac{\varepsilon_{0}\varepsilon_{\rm hBN}}{ed_{\rm top}}(V_{\rm ring}-V^{0}_{\rm ring})$ 
and 
$n_{\rm bulk}=\frac{\varepsilon_{0}\varepsilon_{\rm hBN}}{ed_{\rm bot}}(V_{\rm bg}-V^{0}_{\rm bg})+\frac{\varepsilon_{0}(\varepsilon_{\rm hBN}+\varepsilon_{\rm AlOx})}{e(\varepsilon_{\rm hBN}d_{\rm Alox}+\varepsilon_{\rm AlOx}d_{\rm top})}(V_{\rm tg}-V^{0}_{\rm tg})$.
Here, $\varepsilon_{0}$ is the vacuum permittivity, $\varepsilon_{\rm hBN}$ and $\varepsilon_{\rm AlOx}$ are the dielectric constant of hBN ($\sim$ 3.3) and AlOx ($\sim$ 9.5), $d_{\rm top}$ and $d_{\rm bot}$ are the thickness of the top and bottom hBN, $V^{0}_{\rm bg}$, $V^{0}_{\rm tg}$, and $V^{0}_{\rm ring}$ are the shift of the charge neutrality point from zero.
Similarly, the displacement field inside and outside the ring ($D_{\rm ring}$ and $D_{\rm bulk}$) are expressed as 
$D_{\rm ring}=\frac{\varepsilon_{0}\varepsilon_{\rm hBN}}{ed_{\rm bot}}(V_{\rm bg}-V^{0}_{\rm bg})-\frac{\varepsilon_{0}\varepsilon_{\rm hBN}}{ed_{\rm top}}(V_{\rm ring}-V^{0}_{\rm ring})$
and
$D_{\rm bulk}=\frac{\varepsilon_{0}\varepsilon_{\rm hBN}}{ed_{\rm bot}}(V_{\rm bg}-V^{0}_{\rm bg})-\frac{\varepsilon_{0}(\varepsilon_{\rm hBN}+\varepsilon_{\rm AlOx})}{e(\varepsilon_{\rm hBN}d_{\rm Alox}+\varepsilon_{\rm AlOx}d_{\rm top})}(V_{\rm tg}-V^{0}_{\rm tg})$.
The slope of the charge neutrality lines are attributed to the capacitance ratio between $C_{\rm bg}$ and $C_{\rm ring}$ such as  $-\frac{\partial V_{\rm bg}}{\partial V_{\rm ring}}=\frac{C_{\rm bg}}{C_{\rm ring}}=0.77$ and $-\frac{\partial V_{\rm bg}}{\partial V_{\rm tg}}=\frac{C_{\rm bg}}{C_{\rm tg}}=0.99$.

Figure \ref{gategatemap}(c) shows the two-terminal resistance $R_{xx}$ as functions of magnetic field $B$ and backgate voltage $V_{\rm bg}$. Here, the bulk resistance of the sample is measured by setting $V_{\rm ring}=V_{\rm tg}=0$ V. Clear oscillation of the resistance (so-called Landau fan) is observed, which is attributed to the Shubnikov-de Haas (SdH) oscillation with different carrier densities. The origin of this Landau fan is slightly shifted from zero to $V_{\rm bg}=-0.6$ V, which corresponds to the shift of the charge neutrality point. The behavior of the carrier density determined from the SdH oscillations agrees with the one from the capacitance model within the error of 10 $\%$.

We also measure the four-terminal conductance $G_{\rm 4T}$ using only backgate ($V_{\rm tg}=V_{\rm ring}=0$ V) and estimate the mobility $\mu$ and the mean free path $l_{\rm mfp}$ of the sample as a function of the carrier density $n$. According to the Drude model, it is expected that $G_{\rm 4T}=\frac{\sigma_{xx}Lne\mu}{W}$ and $\sigma_{xx}=\frac{e^{2}}{h}k_{\rm F}l_{\rm mfp}$. Here, $\sigma_{xx}$ is the longitudinal conductivity, $e$ is electron charge, $h$ is Planck constant, $k_{\rm F}=\sqrt{\pi n}$ is the Fermi velocity, $L$ is the length between the source-drain contacts, $W$ is the width between the source and voltage-measurement contacts. We use these relations to estimate the mobility and the mean free path as a function of carrier density obtained from the capacitance model.
Through this measurement, we estimate the mobility $\mu\sim4\times10^{5}$ cm$^2$/Vs for electron ($n>0$) and $\mu\sim2.5\times10^{5}$ cm$^2$/Vs for hole ($n<0$), and the mean free path $l_{\rm mfp}$ from 2.7 $\mu$m (at $n=0.4\times10^{12}$ 1/cm$^2$) to 15 $\mu$m (at $n=2.4\times10^{12}$ 1/cm$^2$). The mean free path $l_{\rm mfp}$ is longer than the sample length ($\pi R\sim1.5$ $\mu$m), which means a ballistic transport through the ring.

\subsection{CARRIER DENSITY AND DISPLACEMENT FIELD DEPENDENCE}
\begin{figure}[H]
\centering
\includegraphics[scale=1.5,pagebox=cropbox,clip]{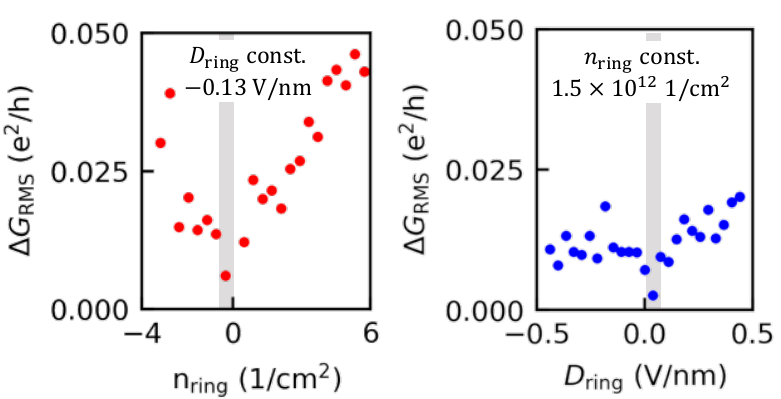}
\caption{$\Delta G_{\rm RMS}$ as a function of (a) $n_{\rm ring}$ with constant $D_{\rm ring}=-0.13$ V/nm and (b) $D_{\rm ring}$ with constant $n_{\rm ring}=1.5\times10^{12}$ /cm$^2$. Gray shaded regions illustrate where the ring becomes ill-defined because of $D_{\rm bulk}=0$.}
\label{NandDdep}
\end{figure}

\hspace{\parindent} To clarify the contribution of carrier density and displacement field to the AB oscillations, we take the $n_{\rm ring}$ and $D_{\rm ring}$ dependence independently, while $n_{\rm bulk}$ stays always zero to form a ring. The sweep directions are illustrated by the arrows in Fig. \ref{gategatemap} (a). Due to this sweep method, the electrostatic definition of the ring becomes imperfect at $n_{\rm ring}\sim 0$ or $D_{\rm ring}\sim 0$ (charge neutrality but ungapped), making less contrast in resistance between the ring and the bulk. The gray shadowed region in Fig. \ref{NandDdep} illustrates such a region.
%

Figure \ref{NandDdep}(a) shows the $n_{\rm ring}$ dependence of the $\Delta G_{\rm RMS}$ with constant $D_{\rm ring}=-0.13$ V/nm.
The oscillation amplitude $\Delta G_{\rm RMS}$ decreases with $n_{\rm ring}$ approaches zero (charge neutrality point). This is the same tendency as the $V_{\rm ring}$ dependence shown in Fig. 2(a) of the main text.
Figure \ref{NandDdep}(b) shows the $D_{\rm ring}$ dependence of the $\Delta G_{\rm RMS}$ with constant $n_{\rm ring}=1.5\times10^{12}$ /cm$^2$. Here, $\Delta G_{\rm RMS}$ varies much less than the case in $n_{\rm ring}$ dependence.
These results suggest that the carrier density inside the ring $n_{\rm ring}$ is the dominant factor to determine the oscillation amplitude.

\subsection{MAGNETIC FIELD DEPENDENCE OF OSCILLATION AMPLITUDE}
\hspace{\parindent} 
%
\begin{figure}[H]
\centering
\includegraphics[scale=1.5,pagebox=cropbox,clip]{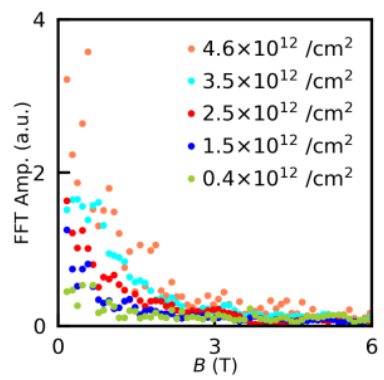}
\caption{The magnetic field dependence of the FFT amplitude at $h/e$ peak. $V_{\rm ring}$ is swept from -5.5, -3.5, -1.5, 0.5, to 2.5 V ($n_{\rm ring}=$ -4.6, -3.5, -2.5, -1.5, and -0.4 cm$^2$).}
\label{Bdep}
\end{figure}

Here, we present the development of the oscillation amplitude ($\frac{h}{e}$ peak) as a function of the magnetic field for all the measured carrier densities ($n_{\rm ring}=$ -4.6, -3.5, -2.5, -1.5, and -0.4 /cm$^2$). Data for $|n_{\rm ring}|=$ 4.6 and 0.4 /cm$^2$ are shown in the inset of Fig. \ref{highmag}(b).
\end{document}